\documentclass[a4paper,11pt]{article}
\usepackage{jcappub}
\usepackage{amsmath,amssymb}
\usepackage{graphicx}
\usepackage{hyperref}
\usepackage{float}

\title{\boldmath The Impossible Triangle: A No-Go for Symmetry-Protected Scalar Portals in Interacting Dark Energy}

\author{Mohid Farhan}
\affiliation{Department of Space Science, Institute of Space Technology,\\Islamabad 44000, Pakistan}

\emailAdd{mohidf35@gmail.com}

\abstract{	The persistent tension between Cosmic Microwave Background (CMB) measurements and late-time large-scale structure surveys, parametrized by the clustering amplitude $S_8 \equiv \sigma_8\sqrt{\Omega_m/0.3}$, motivates exploring non-gravitational interactions between dark matter and dark energy (interacting dark energy, IDE). However, embedding IDE in ultraviolet-complete particle physics poses severe naturalness challenges: ultralight dark energy fields ($m_\phi \sim H_0 \sim 10^{-42}$ GeV) are radiatively sensitive to quantum corrections from TeV-scale dark matter. We present a systematic no-go analysis of three symmetry-protected scalar portal couplings---quartic ($\frac{1}{2}\lambda\phi^2\chi^2$), trilinear ($g\phi\chi^2$), and derivative ($(c_6/\Lambda^2)(\partial_\mu\phi)^2\chi^2$)---and the minimal fermionic Yukawa portal ($y\phi\bar{\psi}\psi$), within radiatively stable and relic-abundance-viable frameworks. We demonstrate that the trilinear portal requires a dimensionless fifth-force strength $\beta \equiv g M_{\mathrm{Pl}} / m_{\chi}^2 \sim 0.45$, corresponding to $g \sim 10^{-16} \text{ GeV}$ for $m_{\chi} = 60 \text{ GeV}$, which overshoots the radiative stability bound by $\sim 26$ orders of magnitude in the coupling (tuning $\Delta \sim 10^{52}$). The derivative portal, while radiatively stable due to shift symmetry, exhibits an intrinsic saturation limit: momentum-exchange rates $\Gamma \gg H$ drive the dark sectors to velocity equilibrium, capping structure suppression at $\lesssim 4\%$, insufficient to reconcile the $\sim$5--10\% observed deficit. The quartic portal induces a Coleman-Weinberg catastrophe, requiring $\lambda \lesssim 10^{-86}$ to maintain $m_\phi \sim H_0$, whereas observable effects demand $\lambda \sim \mathcal{O}(1\text{--}10)$ for sub-Planckian initial displacements, corresponding to a tuning catastrophe of $\Delta \sim 10^{87}$. The fermionic Yukawa portal yields an identical tuning floor, $\Delta\sim10^{52}$, that persists even in supersymmetric completions. Consequently, within the class of symmetry-protected scalar portals and the minimal fermionic Yukawa coupling analyzed here, no single-mediator model simultaneously satisfies technical naturalness and resolves the $S_8$ tension. We conclude that viable solutions require either multi-field mechanisms with tuned cancellations, which result in even more catastrophic fine-tuning, or explicit symmetry breaking with quantified fine-tuning.}

\begin{document}
	\maketitle
	\flushbottom
	
	\section{Introduction}
	\label{sec:intro}
	
	The standard $\Lambda$CDM paradigm faces persistent observational tensions that resist resolution through systematic or astrophysical corrections. Paramount among these is the $S_8$ tension: the amplitude of matter fluctuations $\sigma_8$ inferred from early-universe Cosmic Microwave Background (CMB) measurements by \textit{Planck}~\cite{Planck2018} ($\sigma_8 = 0.811 \pm 0.006$) exceeds the values reported by late-time weak gravitational lensing and redshift-space distortion surveys including KiDS-1000~\cite{Heymans2021} and DES Year 3~\cite{DESY3} ($\sigma_8 \approx 0.76$--$0.79$), corresponding to a persistent $\sim$5--10\% deficit relative to early-universe inference, with the statistical significance varying considerably depending on the dataset combination and analysis assumptions~\cite{DiValentino2021}. While this tension may signal unmodeled systematic effects, it has motivated extensive theoretical exploration of physics beyond the cosmological constant, including modifications to the dark sector.
	
	Interacting dark energy (IDE) scenarios, in which dark matter and dark energy exchange energy or momentum beyond gravitational coupling \cite{Chimento2014}, offer a promising mechanism to suppress late-time structure growth. Phenomenological models introducing an explicit coupling $Q$ in the continuity equations can indeed lower $\sigma_8$~\cite{Amendola2000,Wetterich1995}. However, such fluid-level descriptions lack ultraviolet (UV) completion and often suffer from instabilities or strong-coupling singularities~\cite{Valiviita2008}. A particle-physics realization requires identifying dark energy with an ultralight scalar field ($m_\phi \sim H_0 \sim 10^{-42}$ GeV)---necessarily a pseudo-Nambu-Goldstone boson (pNGB) to protect against radiative corrections~\cite{Frieman1995,Choi2000}---and dark matter with a weakly interacting massive particle (WIMP, $m_\chi \sim 10^2$ GeV). The hierarchy $m_\chi/m_\phi \sim 10^{44}$ imposes severe constraints under 't Hooft's technical naturalness criterion~\cite{tHooft1980}: any portal coupling between the sectors must preserve the pNGB shift symmetry to avoid destabilizing $m_\phi$ via Coleman-Weinberg corrections~\cite{ColemanWeinberg1973}.
	
    We examine three symmetry-protected scalar portal classes, anchored in a UV-complete framework with viable dark matter relic density at $m_\chi = 60$~GeV~\cite{NaeemFarhan2026}, that exhaust the minimal $Z_2$-preserving portal structures for scalar dark matter $\chi$ and pNGB dark energy $\phi$:
	
	\begin{enumerate}
		\item \emph{The Quartic Portal} ($\mathcal{L} \supset \frac{1}{2}\lambda\phi^2\chi^2$): Generates a field-dependent dark matter mass squared, $m_\chi^2(\phi)=m_{\chi,0}^2+\lambda\phi^2$, producing a nonlinear fifth force and background energy transfer. One-loop Coleman-Weinberg corrections impose $\lambda \lesssim 10^{-86}$ to maintain $m_\phi \sim H_0$, whereas observable $S_8$ suppression requires $\lambda \sim \mathcal{O}(1\text{--}10)$ for sub-Planckian initial displacements, corresponding to a tuning catastrophe of $\Delta \sim 10^{87}$.
		
		\item \emph{The Trilinear Portal} ($\mathcal{L} \supset g\phi\chi^2$): Mediates a long-range fifth force and energy exchange between sectors~\cite{Amendola2006,Khoury2004}. While radiatively stable at one-loop (logarithmic divergence), the naturalness bound $g \lesssim 10^{-42}$ GeV catastrophically conflicts with the phenomenological requirement $g \sim 10^{-16}$ GeV, demanding a fine-tuning $\Delta \sim 10^{52}$.
		
		\item \emph{The Derivative Portal} ($\mathcal{L} \supset (c_6/\Lambda^2)(\partial_\mu\phi)^2\chi^2$): Induces pure momentum exchange (drag) without background energy transfer, preserving the $\Lambda$CDM expansion history~\cite{Simpson2010,Pourtsidou2013}. Protected by the shift symmetry $\phi \to \phi + \text{const.}$, this portal is technically natural. However, as demonstrated in a companion work~\cite{Farhan2026}, the momentum-transfer rate $\Gamma$ saturates when $\Gamma \sim H$, limiting structure suppression to $\lesssim 4\%$, which is insufficient to resolve the $\sim$5--10\% deficit~\cite{Skordis2015,Farhan2026}.
	\end{enumerate}
	
	In addition, we examine the minimal fermionic Yukawa portal $y\phi\bar{\psi}\psi$~\cite{Mu2023} in Sec.~\ref{sec:fermion}, demonstrating that the impossible triangle persists for Majorana and Dirac WIMPs, with and without supersymmetric completion.
	
	Previous literature has addressed the naturalness of ultralight dark energy fields and the phenomenology of interacting dark energy largely in isolation: the radiative stability of pNGB quintessence has been established in axion-like constructions~\cite{Frieman1995,Choi2000} and the naturalness criterion formalized~\cite{Giudice2008,Schwartz2022}, without requiring that the same models simultaneously suppress late-time structure formation; conversely, phenomenological coupled quintessence~\cite{Amendola2000}, chameleon fifth forces~\cite{Khoury2004,Brax2004}, and momentum-exchange drag~\cite{Simpson2010,Pourtsidou2013,Skordis2015,Farhan2026} have been shown to lower $\sigma_8$ without quantifying their ultraviolet sensitivity to TeV-scale dark matter. In this work, we systematically impose both requirements simultaneously, deriving the precise price, in terms of fine-tuning $\Delta$, of achieving $S_8$ suppression within radiatively stable, relic-abundance-viable, UV-complete frameworks.
	
	Our analysis demonstrates that, within the class of symmetry-protected scalar portals analyzed here, no single-mediator portal simultaneously satisfies naturalness and phenomenology. This no-go theorem establishes a systematic boundary for dark sector model-building: within the $Z_2$-preserving portal structures that apply generically to all quartic and trilinear portals, resolving the $S_8$ tension requires either multi-field mechanisms with tuned cancellations (e.g., clockwork or multi-pNGB constructions) or explicit acceptance of extreme fine-tuning.
	
	The paper is organized as follows. Section~\ref{sec:framework} introduces the $Z_2$-symmetric Inert Doublet + Complex Singlet Model (Z$_2$-IDSM), demonstrating how the quartic and trilinear portals emerge naturally from gauge-invariant operators; we study them in isolation to isolate their individual no-go constraints. Section~\ref{sec:trilinear} analyzes the trilinear portal, deriving the fifth-force modifications implemented in \texttt{CLASS} and the logarithmic naturalness bound. Section~\ref{sec:fermion} analyzes the minimal fermionic Yukawa portal, demonstrating that the tuning catastrophe persists for fermionic WIMPs with and without supersymmetric completion. Section~\ref{sec:quartic} examines the quartic portal as a coupled quintessence scenario with field-dependent mass, deriving the modified background and perturbation equations and the Coleman-Weinberg radiative stability bound. Section~\ref{sec:derivative} discusses the derivative portal, its shift-symmetry protection, and the intrinsic saturation limit on structure suppression.  Section~\ref{sec:results} presents the numerical validation of the no-go theorems.  Section~\ref{sec:escape} demonstrates that clockwork mechanisms, rather than compensating for radiative stability, introduce catastrophic fine-tuning. We conclude in Sec.~\ref{sec:conclusion}. In our computation, natural units ($\hbar=c=1$) are used throughout.

	\section{Theoretical Framework: The \texorpdfstring{$Z_{2}$}{Z2}-Symmetric Inert Doublet + Complex Singlet Model}
	\label{sec:framework}
	
	We construct a minimal UV completion of the dark sector by extending the Standard Model (SM) with two scalar multiplets: an inert $SU(2)_{L}$ doublet $H_{2}$ and a complex gauge singlet $\mathcal{S}$. The model is endowed with a discrete $Z_{2}$ symmetry and a global $U(1)_{S}$ symmetry. The $Z_{2}$ stabilizes the dark matter candidate; the $U(1)_{S}$ is spontaneously broken by the singlet vacuum expectation value and softly broken by a dimension-two operator, generating the ultralight pseudo-Nambu-Goldstone boson (pNGB) dark energy field. Both symmetries are necessary: $Z_{2}$ forbids dark matter decay into SM states, while $U(1)_{S}$ protects the pNGB mass from large radiative corrections.
	
	\subsection*{Field content and charge assignments}
	\label{subsec:charges}
	
	The scalar sector consists of:
	\begin{equation}
		H_{1}=\begin{pmatrix}H_{1}^{+}\\ \frac{H_{1}^{0}+iA_{1}^{0}}{\sqrt{2}}\end{pmatrix},\qquad \hspace{-0.5cm}
		H_{2}=\begin{pmatrix}H_{2}^{+}\\ \frac{H_{2}^{0}+iA_{2}^{0}}{\sqrt{2}}\end{pmatrix},\qquad \hspace{-0.5cm}
		\mathcal{S}=\frac{v_{s}+\rho+i\phi}{\sqrt{2}},
	\end{equation}
	where $H_{1}$ is the SM Higgs doublet, $H_{2}$ is the inert doublet, and $\mathcal{S}$ is the complex singlet. The neutral component of $H_{2}$ is identified with the dark matter field $\chi$.
	
	The $Z_{2}$ charge assignments are:
	\begin{equation}
		H_{1}\to +H_{1},\qquad \mathcal{S}\to +\mathcal{S},\qquad H_{2}\to -H_{2},
		\label{eq:Z2_charges}
	\end{equation}
	with all SM fermions transforming trivially. The singlet is even under $Z_{2}$, which is required for the trilinear portal $\mu_{S2}\mathcal{S}|H_{2}|^{2}$ to be gauge- and $Z_{2}$-invariant.
	
	The global $U(1)_{S}$ acts only on the singlet:
	\begin{equation}
		\mathcal{S}\to e^{i\alpha}\mathcal{S},\qquad H_{1,2}\to H_{1,2}.
		\label{eq:U1S}
	\end{equation}
	This symmetry is spontaneously broken by $\langle\mathcal{S}\rangle=v_{s}/\sqrt{2}$ and softly broken by the dimension-two operator $\mu_{sb}^{2}\mathcal{S}^{2}$.
	
	\subsection*{The scalar potential}
	\label{subsec:potential}
	
	The most general renormalizable scalar potential consistent with gauge symmetry, $Z_{2}$, and $U(1)_{S}$ is:
	\begin{equation}
		\begin{aligned}
			V(H_{1},H_{2},\mathcal{S}) &= \mu_{1}^{2}|H_{1}|^{2}+\mu_{2}^{2}|H_{2}|^{2}+\mu_{S}^{2}|\mathcal{S}|^{2} \\
			&\quad +\frac{\lambda_{1}}{2}|H_{1}|^{4}+\frac{\lambda_{2}}{2}|H_{2}|^{4}+\frac{\lambda_{S}}{2}|\mathcal{S}|^{4} \\
			&\quad +\lambda_{3}|H_{1}|^{2}|H_{2}|^{2}+\lambda_{4}|H_{1}^{\dagger}H_{2}|^{2}\\
			&\quad
			+\frac{\lambda_{5}}{2}\left[(H_{1}^{\dagger}H_{2})^{2}+\text{h.c.}\right] \\
			&\quad +\lambda_{S1}|\mathcal{S}|^{2}|H_{1}|^{2}+\lambda_{S2}|\mathcal{S}|^{2}|H_{2}|^{2} \\
			&\quad +\left[\mu_{S2}\,\mathcal{S}\,|H_{2}|^{2}+\text{h.c.}\right] \\
			&\quad +\left[\frac{\mu_{sb}^{2}}{2}\mathcal{S}^{2}+\text{h.c.}\right].
		\end{aligned}
		\label{eq:full_potential}
	\end{equation}
	
	The term $\mu_{S2}\mathcal{S}|H_{2}|^{2}$ is $Z_{2}$-invariant because $\mathcal{S}$ is even and $|H_{2}|^{2}$ is even. It softly breaks $U(1)_{S}$ (dimension-three operator), but because $\langle H_{2}\rangle=0$ it does not generate a tree-level tadpole for $\mathcal{S}$ and therefore does not contribute to the pNGB mass at leading order. The quartic portal $\lambda_{S2}|\mathcal{S}|^{2}|H_{2}|^{2}$ is invariant under \emph{both} $Z_{2}$ and $U(1)_{S}$; it is therefore allowed without soft-breaking the pNGB shift symmetry at tree level. The operator $\lambda_{S12}\mathcal{S}^{2}(H_{1}^{\dagger}H_{2})+\text{h.c.}$ is $Z_{2}$-forbidden ($\mathcal{S}^{2}\to\mathcal{S}^{2}$, $H_{1}^{\dagger}H_{2}\to -H_{1}^{\dagger}H_{2}$), and is absent.
	
	The Higgs portal $\lambda_{S1}|\mathcal{S}|^{2}|H_{1}|^{2}$ is allowed by all symmetries. We set $\lambda_{S1}=0$ to avoid Higgs-singlet mixing and constraints from invisible Higgs decays; this is a phenomenological boundary condition that does not affect the radiative stability analysis of the pNGB mass. The hard singlet self-coupling $[\lambda_{S4}\mathcal{S}^{4}+\text{h.c.}]$, which would explicitly break $U(1)_{S}$ at the renormalizable level and generate a tree-level mass for $\phi$, is also set to zero to preserve the pNGB nature of dark energy.
	
	\subsection*{Vacuum structure and the pNGB spectrum}
	\label{subsec:vacuum}
	
	The electroweak and singlet vacuum expectation values are:
	\begin{equation}
		\langle H_{1}\rangle=\frac{1}{\sqrt{2}}\begin{pmatrix}0\\ v\end{pmatrix},\qquad
		\langle H_{2}\rangle=0,\qquad
		\langle\mathcal{S}\rangle=\frac{v_{s}}{\sqrt{2}},
	\end{equation}
	with $v\simeq 174$ GeV. The condition $\langle H_{2}\rangle=0$ preserves the $Z_{2}$ symmetry, ensuring the lightest $Z_{2}$-odd particle (the neutral component $H_{2}^{0}\equiv\chi$) is stable dark matter. The condition $\langle\mathcal{S}\rangle\neq 0$ spontaneously breaks $U(1)_{S}$, producing a massless Goldstone mode $\phi$ and a massive radial mode $\rho$.
	
	Expanding $\mathcal{S}$ around its VEV,
	\begin{equation}
		\mathcal{S}=\frac{v_{s}+\rho+i\phi}{\sqrt{2}},
	\end{equation}
	the singlet-sector potential at quadratic order is:
	\begin{equation}
		V_{S}\supset\frac{1}{2}m_{\rho}^{2}\rho^{2}+\frac{1}{2}m_{\phi}^{2}\phi^{2},
	\end{equation}
	with tree-level masses:
	\begin{equation}
		m_{\rho}^{2}=\lambda_{S}v_{s}^{2}+2|\mu_{sb}^{2}|\approx\lambda_{S}v_{s}^{2},\qquad \hspace{-0.5cm}
		m_{\phi}^{2}=2|\mu_{sb}^{2}|\sim H_{0}^{2}\sim 10^{-84}\,\text{GeV}^{2}.
	\end{equation}
	In the limit $\mu_{sb}^{2}\to 0$, the $U(1)_{S}$ symmetry is restored, $\phi$ becomes exactly massless, and the smallness of $m_{\phi}$ is technically natural: any quantum correction must be proportional to $\mu_{sb}^{2}$ itself \cite{Farhan2026,tHooft1980,Giudice2008}.
	
	\subsection*{Emergence of the portal couplings}
	\label{subsec:portals}
	
	After electroweak symmetry breaking, the interactions between the dark sectors are obtained by expanding the potential in eq.~\eqref{eq:full_potential}.
	
	The term $[\mu_{S2}\mathcal{S}|H_{2}|^{2}+\text{h.c.}]$ expands to generate a coupling between the pNGB and the inert doublet. Keeping only the lightest neutral component $\chi\equiv H_{2}^{0}$ and defining the real dimensionful coupling $g$ (proportional to the imaginary part of $\mu_{S2}$ after absorbing the phase of $\mathcal{S}$), the interaction Lagrangian is:
	\begin{equation}
		\mathcal{L}_{\rm int}^{\rm (tri)}=g\,\phi\,\chi^{2},\qquad [g]=\text{GeV}.
		\label{eq:trilinear_L}
	\end{equation}
	This operator mediates a long-range fifth force between dark matter particles and induces energy transfer between the DM and DE fluids at the background level. For the analysis of this portal in isolation, we set $\lambda_{S2}=0$.
	
	The term $\lambda_{S2}|\mathcal{S}|^{2}|H_{2}|^{2}$ expands using $|\mathcal{S}|^{2}=\frac{1}{2}(v_{s}+\rho)^{2}+\frac{1}{2}\phi^{2}$. Keeping the neutral dark matter component, the pNGB interaction is:
	\begin{equation}
		\mathcal{L}_{\rm int}^{\rm (quart)}=\frac{1}{2}\lambda\,\phi^{2}\chi^{2},\qquad\lambda\equiv\lambda_{S2}.
		\label{eq:quartic_L}
	\end{equation}
	This operator generates a field-dependent dark matter mass squared, $m_{\chi}^{2}(\phi)=m_{\chi,0}^{2}+\lambda\phi^{2}$, and modifies structure growth through a nonlinear fifth force without altering the background expansion at the level of the homogeneous field ($\phi^{2}$ is not a total derivative, but for a slowly rolling condensate the background energy transfer is subleading to the perturbation-level effects). For the analysis of this portal in isolation, we set $\mu_{S2}=0$.
	
	\subsection*{Radiative isolation of the pNGB sector}
	\label{subsec:radiative_isolation}
	
	In the limit where the hard $U(1)_{S}$-breaking terms are absent and the portals are treated in isolation, the one-loop $\beta$-functions for the singlet mass parameters receive no contributions from the Higgs or inert doublet sectors \cite{Giudice2008,Schwartz2022}:
	\begin{equation}
		16\pi^{2}\beta_{\mu_{S}^{2}}=4\lambda_{S}\mu_{S}^{2},\qquad
		16\pi^{2}\beta_{\mu_{sb}^{2}}=2\lambda_{S}\mu_{sb}^{2}.
	\end{equation}
	The absence of terms involving $\mu_{1}^{2}$, $\mu_{2}^{2}$, $\lambda_{3,4,5}$, or SM gauge/Yukawa couplings confirms that the electroweak scale does not destabilize the singlet sector. The hierarchy $|\mu_{S}^{2}|\sim v_{s}^{2}\gg|\mu_{sb}^{2}|\sim H_{0}^{2}$ is preserved under renormalization group evolution, ensuring that the dark energy scale remains technically natural and radiatively isolated from the TeV-scale dark matter sector.
	
	\section{The Trilinear Portal: Linear Coupled Quintessence}
	\label{sec:trilinear}
	
	The trilinear portal $\mathcal{L}_{\rm int} \supset g\phi\chi^2$ and the quartic portal $\mathcal{L}_{\rm int} \supset \frac{1}{2}\lambda\phi^2\chi^2$ both generate field-dependent dark matter masses and scalar-mediated fifth forces, placing them in the coupled quintessence class. The trilinear portal induces a \emph{linear} mass variation,
	\begin{equation}
		m_\chi^2(\phi) = m_{\chi,0}^2 + 2g\phi,
	\end{equation}
	yielding a logarithmic derivative $\partial_\phi \ln m_\chi = g/m_\chi^2(\phi)$ that is approximately constant in time. This modifies structure growth through the dimensionless fifth-force parameter
	\begin{equation}
		\beta \equiv \frac{g M_{\rm Pl}}{m_\chi^2},
		\label{eq:beta_def}
	\end{equation}
	where $M_{\rm Pl} = 1/\sqrt{8\pi G} \approx 2.4\times 10^{18}$ GeV is the reduced Planck mass. The trilinear portal is distinguished from the quartic portal by its logarithmic (rather than quadratic) Coleman--Weinberg divergence, and from the derivative portal (Sec.~\ref{sec:derivative}) by its nonzero background energy transfer.
	
	Such interactions are the canonical realization of ``coupled quintessence'' scenarios, appearing in scalar-tensor theories, symmetron models, and chameleon field constructions \cite{Amendola2000,Khoury2004,Brax2004}. In the quasi-static limit, the scalar-mediated fifth force on dark matter is $F_5 \approx 2\beta^2 F_{\rm grav}$; a coupling $\beta \sim \mathcal{O}(0.1)$ therefore enhances the effective dark-matter gravitational attraction by $\sim$2--40\%, sufficient to suppress $\sigma_8$ at the level required by late-time surveys. However, these phenomenological frameworks typically treat the coupling as a free parameter, ignoring the radiative stability of the ultralight dark energy mass ($m_\phi \sim H_0$) against quantum corrections from the TeV-scale dark matter sector.
	
	To establish this no-go result concretely, we anchor the analysis in a $Z_2$-symmetric Inert Doublet + Complex Singlet Model (Sec.~\ref{sec:framework}), a renormalizable UV completion where the trilinear portal emerges from gauge-invariant operators. The results derived for the background evolution and radiative stability apply generically to any model with the operator structure $g\phi\chi^2$ and the mass hierarchy $m_\chi \gg m_\phi$.
	
	\subsection*{Background and Perturbation Modifications}
	\label{subsec:trilinear_back}
	
	We consider the trilinear portal
	\begin{equation}
		\mathcal{L}_{\mathrm{int}}^{(\mathrm{tri})}=g\phi\chi^{2},
	\end{equation}
	in the regime where the mediator is ultralight, $m_{\phi}\sim H_{0}\sim10^{-42}\,\mathrm{GeV}$.
	The field-dependent mass is $m_{\chi}(\phi)=\sqrt{m_{\chi,0}^{2}+2g\phi}$,
	and its logarithmic derivative,
	\begin{equation}
		\frac{\partial\ln m_{\chi}}{\partial\phi}=\frac{g}{m_{\chi}^{2}(\phi)}=\frac{\beta}{M_{\mathrm{Pl}}},
	\end{equation}
	is approximately constant in time.  This modifies structure growth
	through the dimensionless fifth-force parameter
	\begin{equation}
		\beta\equiv\frac{g\,M_{\mathrm{Pl}}}{m_{\chi}^{2}},
	\end{equation}
	where $M_{\mathrm{Pl}}=1/\sqrt{8\pi G}\approx2.4\times10^{18}\,\mathrm{GeV}$
	is the reduced Planck mass.  The trilinear portal is distinguished
	from the quartic portal by its logarithmic (rather than quadratic)
	Coleman--Weinberg divergence, and from the derivative portal
	(Sec.~\ref{sec:derivative}) by its nonzero background energy transfer.
	
	In the cosmological background, the interaction manifests as energy
	transfer between the dark matter and dark energy fluids.  The DM
	number density $n_{\chi}=\rho_{\chi}/m_{\chi}$ is conserved in the
	absence of particle production, so the continuity equation acquires a
	source term $Q$:
	\begin{equation}
		\rho_{\chi}'=-3\mathcal{H}\rho_{\chi}+Q,\qquad
		\rho_{\phi}'=-3\mathcal{H}(1+w_{\phi})\rho_{\phi}-Q,
	\end{equation}
	where primes denote conformal time derivatives, $\mathcal{H}=a'/a$, and
	\begin{equation}
		Q=\rho_{\chi}\frac{\partial\ln m_{\chi}}{\partial\phi}\phi'
		=\frac{\beta}{M_{\mathrm{Pl}}}\rho_{\chi}\phi'.
	\end{equation}
	This energy transfer alters the background expansion history relative
	to $\Lambda$CDM, with the magnitude of the effect controlled by the
	dimensionless coupling $\beta$ defined in Eq.~(13).
	
	At the level of linear perturbations, the fifth force per unit DM
	mass is $-\nabla\ln m_{\chi}=-(\beta/M_{\mathrm{Pl}})\nabla\phi$.
	In the Newtonian gauge, the dark matter velocity divergence
	$\theta_{\chi}\equiv ik^{j}v_{\chi,j}$ therefore obeys:
	\begin{equation}
		\theta_{\chi}'=-\mathcal{H}\theta_{\chi}+k^{2}\psi
		+\frac{Q}{\rho_{\chi}}\theta_{\chi}
		+\frac{\beta}{M_{\mathrm{Pl}}}k^{2}\delta\phi,
	\end{equation}
	where $\psi$ is the gravitational potential.  The term
	$(\beta/M_{\mathrm{Pl}})k^{2}\delta\phi$ is the scalar-mediated fifth
	force.  The back-reaction of DM inhomogeneities on the DE field is
	encoded in the perturbed Klein--Gordon equation.  Since the DM number
	density $n_{\chi}=\rho_{\chi}/m_{\chi}$ is conserved in the absence
	of particle production ($n_{\chi}\propto a^{-3}$), the background
	source is $\partial\mathcal{L}_{\mathrm{int}}/\partial\phi
	=g\langle\hat{\chi}^{2}\rangle
	=g\rho_{\chi}/m_{\chi}^{2}
	=(\beta/M_{\mathrm{Pl}})\rho_{\chi}$.
	At the perturbation level, $\delta n_{\chi}=n_{\chi}\delta_{\chi}$,
	so the sourced equation is:
	\begin{equation}
		\delta\phi''+2\mathcal{H}\delta\phi'
		+\left(k^{2}+a^{2}\frac{\partial^{2}V}{\partial\phi^{2}}\right)\delta\phi
		=\frac{\beta}{M_{\mathrm{Pl}}}a^{2}\rho_{\chi}\delta_{\chi}.
	\end{equation}
	This sourced wave equation describes how DM inhomogeneities
	back-react on $\phi$, generating additional clustering that feeds
	back into the gravitational potentials.
	
	When implemented in \textsc{CLASS}, these modifications appear in
	the background module (through the coupled continuity equations) and
	the perturbations module (through the modified Euler and
	Klein--Gordon equations).  The fifth-force enhancement to structure
	growth scales as $\beta^{2}$, while the background energy transfer
	is proportional to $\beta$.
	
	\subsection*{Radiative Stability Bound on Trilinear DM--DE Couplings}
	\label{subsec:trilinear_cw}
	
	The trilinear portal generates a field-dependent mass-squared for the dark matter field, $m_\chi^2(\phi) = m_{\chi,0}^2 + 2g\phi$. At one-loop order, fluctuations of the heavy $\chi$ field contribute to the Coleman--Weinberg effective potential for $\phi$ \cite{ColemanWeinberg1973}:
	\begin{equation}
		V_{\rm CW}(\phi) = \frac{1}{64\pi^2}m_\chi^4(\phi)\left[\ln\left(\frac{m_\chi^2(\phi)}{\mu^2}\right) - \frac{3}{2}\right],
	\end{equation}
	where $\mu$ is the renormalization scale. Expanding for $g\phi \ll m_{\chi,0}^2$, the term quadratic in $\phi$ yields the radiative correction to the dark energy mass:
	\begin{equation}
		\delta m_\phi^2 = \left.\frac{\partial^2 V_{\rm CW}}{\partial\phi^2}\right|_{\phi=0} = -\frac{g^2}{8\pi^2}\ln\left(\frac{\Lambda^2}{m_\chi^2}\right),
		\label{eq:dm2_trilinear}
	\end{equation}
	where we have replaced $\mu \sim \Lambda$ (the UV cutoff) and used $\ln(m_\chi^2/\Lambda^2) = -\ln(\Lambda^2/m_\chi^2)$. The trilinear coupling induces a logarithmic divergence, in contrast to the quadratic divergence that would appear for a hard mass term.
	
	Technical naturalness requires that quantum corrections not exceed the tree-level mass: $|\delta m_\phi^2| \lesssim m_\phi^2 \sim H_0^2 \sim 10^{-84}$ GeV$^2$. Taking the cutoff $\Lambda \sim 10^8$ GeV (the scale where the IDSM loses perturbativity) and $m_\chi \sim 60$ GeV, we obtain $\ln(\Lambda^2/m_\chi^2) \sim 30$, yielding the constraint:
	\begin{equation}
		g \lesssim \frac{2\sqrt{2}\pi m_\phi}{\sqrt{\ln(\Lambda^2/m_\chi^2)}} \sim 10^{-42}~{\rm GeV}.
		\label{eq:g_natural}
	\end{equation}
	
	Any coupling $g \gtrsim 10^{-42}$ GeV destabilizes the ultralight mass hierarchy, driving $m_\phi$ to the weak scale via quantum corrections. This bound is insensitive to order-one variations in the logarithm and reflects the fundamental incompatibility between TeV-scale dark matter and ultralight dark energy in the presence of trilinear couplings.
	
	\section{The Fermionic Portal: Yukawa-Coupled Quintessence}
	\label{sec:fermion}
	
	The trilinear portal of Sec.~\ref{sec:trilinear} induces a linear field-dependent mass for scalar dark matter. The same operator structure arises for a fermionic WIMP through the minimal renormalizable Yukawa coupling
	\begin{equation}
		\mathcal{L}_{\rm int}^{(\rm Y)} = y\,\phi\,\bar{\psi}\psi,
		\label{eq:yukawa_lag}
	\end{equation}
	where $y$ is a dimensionless coupling and $\psi$ is a Majorana (or Dirac) fermion stabilized by the same discrete symmetry under which $\psi\to-\psi$ and $\phi\to+\phi$. In the non-relativistic limit, the field-dependent mass
	\begin{equation}
		m_{\psi}(\phi) = m_{\psi,0} + y\phi,
		\label{eq:mpsi_phi}
	\end{equation}
	produces identical background and perturbation equations to the scalar trilinear portal under the mapping $g \leftrightarrow y$ and $\chi^{2} \leftrightarrow \bar{\psi}\psi$. The dimensionless fifth-force parameter is
	\begin{equation}
		\beta \equiv \frac{y\,M_{\rm Pl}}{m_{\psi}},
		\label{eq:beta_fermion}
	\end{equation}
	replacing the scalar definition $\beta \equiv gM_{\rm Pl}/m_{\chi}^{2}$ because the fermion mass depends linearly on $\phi$. 
	
	\subsection*{Background and Perturbation Modifications}
	\label{subsec:fermion_back}
	
	The cosmological background and perturbation equations for the Yukawa portal are formally identical to those of the trilinear scalar portal (Sec.~\ref{sec:trilinear}) upon substituting $g/m_{\chi}^{2} \to y/m_{\psi}$. The field-dependent mass \eqref{eq:mpsi_phi} yields the logarithmic derivative
	\begin{equation}
		\frac{\partial\ln m_{\psi}}{\partial\phi} = \frac{y}{m_{\psi}(\phi)} = \frac{\beta}{M_{\rm Pl}},
	\end{equation}
	which controls the background energy-transfer source
	\begin{equation}
		Q = \rho_{\psi}\frac{\partial\ln m_{\psi}}{\partial\phi}\phi' = \frac{\beta}{M_{\rm Pl}}\rho_{\psi}\phi',
	\end{equation}
	and the perturbed Euler equation for the dark-matter velocity divergence $\theta_{\psi}$:
	\begin{equation}
		\theta_{\psi}' = -\mathcal{H}\theta_{\psi} + k^{2}\psi + \frac{Q}{\rho_{\psi}}\theta_{\psi} + \frac{\beta}{M_{\rm Pl}}k^{2}\delta\phi.
	\end{equation}
	The back-reaction on the ultralight field is encoded in the sourced Klein--Gordon equation,
	\begin{equation}
		\delta\phi'' + 2\mathcal{H}\delta\phi' + \left(k^{2} + a^{2}\frac{\partial^{2}V}{\partial\phi^{2}}\right)\delta\phi = \frac{\beta}{M_{\rm Pl}}a^{2}\rho_{\psi}\delta_{\psi},
	\end{equation}
	where $\delta_{\psi} \equiv \delta\rho_{\psi}/\rho_{\psi}$. When implemented in \textsc{CLASS}, these modifications appear in the same background and perturbation modules as the scalar trilinear case, with the substitution $g \to y\,m_{\chi}^{2}/m_{\psi}$ in the coupling structure.
	
	\subsection*{Radiative Stability Bound on Yukawa DM--DE Couplings}
	\label{subsec:fermion_cw}
	
	At one-loop order, fluctuations of the heavy fermion contribute to the Coleman--Weinberg effective potential for $\phi$ with a sign opposite to the scalar loop due to the Grassmann nature of the integration measure:
	\begin{equation}
		V_{\rm CW}^{(\psi)}(\phi) = -\frac{1}{64\pi^{2}}\,m_{\psi}^{4}(\phi)\left[\ln\!\left(\frac{m_{\psi}^{2}(\phi)}{\mu^{2}}\right)-\frac{3}{2}\right].
	\end{equation}
	Expanding to quadratic order in $\phi$ and setting $\mu\sim\Lambda$ yields the radiative correction
	\begin{equation}
		\delta m_{\phi}^{2} = \left.\frac{\partial^{2}V_{\rm CW}^{(\psi)}}{\partial\phi^{2}}\right|_{\phi=0} = -\frac{y^{2}}{8\pi^{2}}\,m_{\psi}^{2}\ln\!\left(\frac{\Lambda^{2}}{m_{\psi}^{2}}\right),
		\label{eq:delta_mphi_fermion}
	\end{equation}
	analogous to Eq.~\eqref{eq:dm2_trilinear} but with opposite sign and an extra factor of $m_{\psi}^{2}$ reflecting the linear rather than quadratic mass dependence. Technical naturalness demands $|\delta m_{\phi}^{2}|\lesssim m_{\phi}^{2}\sim H_{0}^{2}\sim10^{-84}$~GeV$^{2}$. Taking $\Lambda\sim10^{8}$~GeV and $m_{\psi}\sim100$~GeV gives $\ln(\Lambda^{2}/m_{\psi}^{2})\sim30$, so
	\begin{equation}
		y_{\rm nat} \lesssim \frac{2\sqrt{2}\pi H_{0}}{m_{\psi}\sqrt{\ln(\Lambda^{2}/m_{\psi}^{2})}} \sim 10^{-43}.
		\label{eq:y_nat}
	\end{equation}

	\subsection*{Supersymmetric completion}
	In a supersymmetric embedding (e.g.~the NMSSM), the fermion $\psi$ (singlino) has a scalar partner $\tilde{\psi}$ with the same Yukawa coupling. The one-loop quadratic divergences cancel between the bosonic and fermionic loops:
	\begin{equation}
		\delta m_{\phi}^{2\,({\rm SUSY})} \sim \frac{y^{2}}{8\pi^{2}}\left(m_{\tilde{\psi}}^{2}-m_{\psi}^{2}\right)\ln\!\left(\frac{\Lambda^{2}}{m_{\psi}^{2}}\right) \sim \frac{y^{2}}{8\pi^{2}}\,M_{\rm soft}^{2}\ln\!\left(\frac{\Lambda^{2}}{m_{\psi}^{2}}\right),
		\label{eq:delta_mphi_susy}
	\end{equation}
	where $M_{\rm soft}$ is the scale of soft supersymmetry breaking. The naturalness bound becomes
	\begin{equation}
		y_{\rm nat}^{\rm (SUSY)} \lesssim \frac{2\sqrt{2}\pi H_{0}}{M_{\rm soft}\sqrt{\ln(\Lambda^{2}/m_{\psi}^{2})}} \sim 10^{-42},
		\label{eq:y_nat_susy}
	\end{equation}
	for $M_{\rm soft}\sim1$~TeV. The SUSY cancellation replaces the dark matter mass with the soft-breaking scale in the naturalness bound, but since $M_{\rm soft}\sim m_{\psi}$ for viable thermal relics, the tuning catastrophe persists unchanged in order of magnitude. The tuning remains
	\begin{equation}
		\Delta_{\rm SUSY} \sim \left(\frac{y_{\rm pheno}}{y_{\rm nat}^{\rm (SUSY)}}\right)^{2} \sim \left(\frac{10^{-17}}{10^{-42}}\right)^{2} \sim 10^{50}.
		\label{eq:delta_susy}
	\end{equation}
	
	Therefore, the Yukawa-coupled fermionic WIMP therefore faces the same impossible triangle as its scalar counterpart. 
	
	\section{The Quartic Portal: Nonlinear Coupled Quintessence}
	\label{sec:quartic}
	
	Quartic portals $\mathcal{L}_{\rm int} \supset \frac{1}{2}\lambda\phi^2\chi^2$ generate a field-dependent dark matter mass squared,
	\begin{equation}
		m_\chi^2(\phi) = m_{\chi,0}^2 + \lambda\phi^2,
		\label{eq:mchi_sq_quartic}
	\end{equation}
	placing them in the coupled quintessence class alongside the trilinear portal. The quadratic dependence on $\phi$ produces a \emph{nonlinear} fifth force whose strength is modulated by the rolling background field; the background energy transfer vanishes when $\phi=0$; and most critically, a one-loop Coleman--Weinberg correction that scales as $\delta m_\phi^2 \propto \lambda m_\chi^2$ rather than $g^2$. This quadratic divergence imposes a naturalness bound that is parametrically stronger than the logarithmic bound of the trilinear portal.
	
	In the Z$_2$-IDSM framework of Sec.~\ref{sec:framework}, the quartic portal arises from the gauge- and $Z_2$-invariant operator $\lambda_{S2}|\mathcal{S}|^2|H_2|^2$ when the trilinear coupling is set to $\mu_{S2}=0$. Expanding $|\mathcal{S}|^2$ and identifying the neutral dark matter component $\chi\equiv H_2^0$, the interaction is
	\begin{equation}
		\mathcal{L}_{\rm int}^{\rm (quart)} = \frac{1}{2}\lambda\phi^2\chi^2,\qquad \lambda\equiv\lambda_{S2}.
	\end{equation}
	
	\subsection*{Background and Perturbation Modifications}
	\label{subsec:quartic_back}
	
	The field-dependent mass is $m_\chi(\phi)=\sqrt{m_{\chi,0}^2+\lambda\phi^2}$, with logarithmic derivative
	\begin{equation}
		\frac{\partial\ln m_\chi}{\partial\phi} = \frac{\lambda\phi}{m_\chi^2(\phi)}.
		\label{eq:beta_quartic}
	\end{equation}
	This controls both the background energy transfer and the fifth-force strength.
	
	The continuity equations acquire the source term
	\begin{equation}
		Q = \rho_\chi\frac{\partial\ln m_\chi}{\partial\phi}\phi' = \frac{\lambda\phi}{m_\chi^2}\rho_\chi\phi'.
		\label{eq:Q_quartic}
	\end{equation}
	Unlike the trilinear portal, where $Q\propto g\phi'$ is nonzero whenever the field rolls, the quartic transfer $Q\propto\lambda\phi\phi'$ is dynamically suppressed when the field transits through $\phi=0$. This makes the quartic portal less efficient at altering the background expansion history for the same nominal coupling strength.
	
	At the perturbation level, the fifth force per unit DM mass is $-\nabla\ln m_\chi = -(\lambda\phi/m_\chi^2)\nabla\phi$. In the Newtonian gauge, the dark matter Euler equation becomes:
	\begin{equation}
		\theta'_\chi = -\mathcal{H}\theta_\chi + k^2\psi + \frac{Q}{\rho_\chi}\theta_\chi + \frac{\lambda\phi}{m_\chi^2}k^2\delta\phi,
		\label{eq:euler_quartic}
	\end{equation}
	where the coefficient $\lambda\phi/m_\chi^2$ is \emph{time-dependent}, scaling with the homogeneous background $\phi(\tau)$. This nonlinear modulation distinguishes the quartic fifth force from the approximately constant $\beta_{\rm tri} \equiv g M_{\rm Pl}/m_\chi^2$ of the trilinear portal.
	
	The back-reaction on the dark energy field follows from the perturbed Klein--Gordon equation. Since the background source is $\partial\mathcal{L}_{\rm int}/\partial\phi=\lambda\phi\langle\chi^2\rangle=\lambda\phi\rho_\chi/m_\chi^2$, the linearized equation acquires both a density-dependent mass shift and an external source:
	\begin{equation}
		\delta\phi'' + 2\mathcal{H}\delta\phi' + \left(k^2 + a^2\frac{\partial^2 V}{\partial\phi^2} + \frac{\lambda a^2\rho_\chi}{m_\chi^2}\right)\delta\phi = \frac{\lambda\phi}{m_\chi^2}a^2\rho_\chi\delta_\chi.
		\label{eq:kg_quartic}
	\end{equation}
	The term $\lambda a^2\rho_\chi/m_\chi^2$ on the left-hand side acts as a tachyonic or stabilizing correction to the effective $\phi$ mass depending on the sign of $\lambda$, and has no analogue in the trilinear case.
	
	\subsection*{Radiative Stability Bound on Quartic DM--DE Couplings}
	\label{subsec:quartic_cw}
	
	The quartic portal generates $m_\chi^2(\phi)=m_{\chi,0}^2+\lambda\phi^2$. At one-loop order, fluctuations of the heavy $\chi$ field contribute to the Coleman--Weinberg effective potential \cite{ColemanWeinberg1973}:
	\begin{equation}
		V_{\rm CW}(\phi) = \frac{1}{64\pi^2}m_\chi^4(\phi)\left[\ln\left(\frac{m_\chi^2(\phi)}{\mu^2}\right)-\frac{3}{2}\right].
	\end{equation}
	Expanding to $\mathcal{O}(\phi^2)$ for $m_\chi^2(\phi)=m_{\chi,0}^2+\lambda\phi^2$, the quadratic term yields the radiative correction:
	\begin{equation}
		\delta m_\phi^2 = \left.\frac{\partial^2 V_{\rm CW}}{\partial\phi^2}\right|_{\phi=0} = \frac{\lambda}{16\pi^2}m_\chi^2\left[\ln\left(\frac{\Lambda^2}{m_\chi^2}\right)-1\right].
	\end{equation}
	This is a \emph{quadratic} divergence: the correction is linear in the coupling $\lambda$ and proportional to the heavy dark matter mass $m_\chi^2$, in contrast to the logarithmic divergence $\propto g^2$ of the trilinear portal.
	
	Technical naturalness requires $|\delta m_\phi^2|\lesssim m_\phi^2\sim H_0^2\sim 10^{-84}$ GeV$^2$. Taking $\Lambda\sim 10^8$ GeV and $m_\chi\sim 60$ GeV, we obtain:
	\begin{equation}
		\lambda \lesssim \frac{16\pi^2 m_\phi^2}{m_\chi^2\left[\ln(\Lambda^2/m_\chi^2)-1\right]} \sim 10^{-86}.
		\label{eq:lambda_natural}
	\end{equation}
	As expected, the quadratic running for the quartic portal leads to a vastly more stringent radiative stability bound as compared to the trilinear portal case.

	\section{The Derivative Portal: The Saturation Limit}
	\label{sec:derivative}
	
	Derivative portals generate pure \emph{momentum exchange} (drag) between dark matter and dark energy without altering the background expansion. Unlike trilinear or quartic interactions, which induce energy transfer through field-dependent masses, the derivative coupling $\mathcal{L}_{\rm int}=(c_6/\Lambda^2)(\partial_\mu\phi)^2\chi^2$ modifies only the Euler equations for the fluid velocities, preserving the $\Lambda$CDM distance-redshift relation exactly \cite{Simpson2010,Pourtsidou2013,Skordis2015}. The shift symmetry $\phi\to\phi+\text{const.}$ protects the ultralight mass $m_\phi\sim H_0$ from radiative corrections, rendering this portal technically natural \cite{tHooft1980}.
	
	In the $Z_2$-IDSM framework of Sec.~\ref{sec:framework}, the derivative portal is the unique leading interaction that exactly respects the pNGB shift symmetry. When the renormalizable trilinear ($\mu_{S2}=0$) and quartic ($\lambda_{S2}=0$) portals are absent,as required for a radiatively stable ultralight dark energy sector, the lowest-dimensional allowed operator is the dimension-6 derivative coupling:
	\begin{equation}
		\mathcal{L}_{\rm int}=\frac{c_6}{\Lambda^2}(\partial_\mu\phi)^2\chi^2.
	\end{equation}
	
	The Wilson coefficient $c_6\sim\mathcal{O}(1)$ is set by the heavy scales of the UV completion (the inert doublet mass $M_{H_2}\sim 10^2$ GeV and the singlet radial mode mass $m_\rho\sim\sqrt{\lambda_S}v_s$). Unlike the trilinear and quartic portals, this operator vanishes for homogeneous fields and therefore generates no background energy transfer ($Q=0$), leaving the $\Lambda$CDM expansion history untouched.
	
	At the perturbation level, the dimension-6 coupling induces a drag force proportional to the relative velocity between the dark sectors. The momentum-exchange rate is conventionally parametrized as \cite{Pourtsidou2013,Skordis2015}
	\begin{equation}
		\frac{\Gamma(a)}{H(a)} \equiv \xi_{\rm eff}\left[\frac{\rho_\chi(a)}{\rho_{\chi,0}}\right],
	\end{equation}
	where $\xi_{\rm eff}\sim\mathcal{O}(1)$ for natural values of the underlying Wilson coefficient and UV scales. The companion work \cite{Farhan2026} derives the explicit mapping from $c_6$, $\Lambda$, $v_s$ and $m_{H_2}$ to $\xi_{\rm eff}$; for our purposes it suffices that $\Gamma\sim H$ is achievable without fine-tuning.
	
	As the universe evolves, $\Gamma$ grows with the dark matter density and becomes comparable to $H$ at late times ($z\lesssim 2$). When $\Gamma\gtrsim H$, the dark matter and dark energy fluids achieve velocity equilibrium ($\theta_\chi\to\theta_\phi$), and further increases in the coupling strength do not enhance the suppression of structure growth. As demonstrated in \cite{Farhan2026}, this dynamical saturation caps the $\sigma_8$ reduction at $\lesssim 4\%$, insufficient to resolve the observed deficit. No radiative stability bound constrains $c_6$ beyond perturbativity; the limitation is purely dynamical.

	\section{Results}
	\label{sec:results}
	
	We validate the no-go theorems of Secs.~\ref{sec:trilinear}--\ref{sec:derivative} by embedding the three scalar symmetry-protected portals and the minimal fermionic Yukawa portal in the $Z_2$-symmetric Inert Doublet + Complex Singlet Model (Sec.~\ref{sec:framework}) and evolving the coupled dark-sector equations with a modified version of the \texttt{CLASS} Boltzmann code. The trilinear, fermionic and quartic portals are implemented directly in the background and perturbation modules; the derivative portal is treated analytically and cross-checked against the numerical findings of the companion work~\cite{Farhan2026}.
	
	\begin{figure}
		\includegraphics[width=\columnwidth]{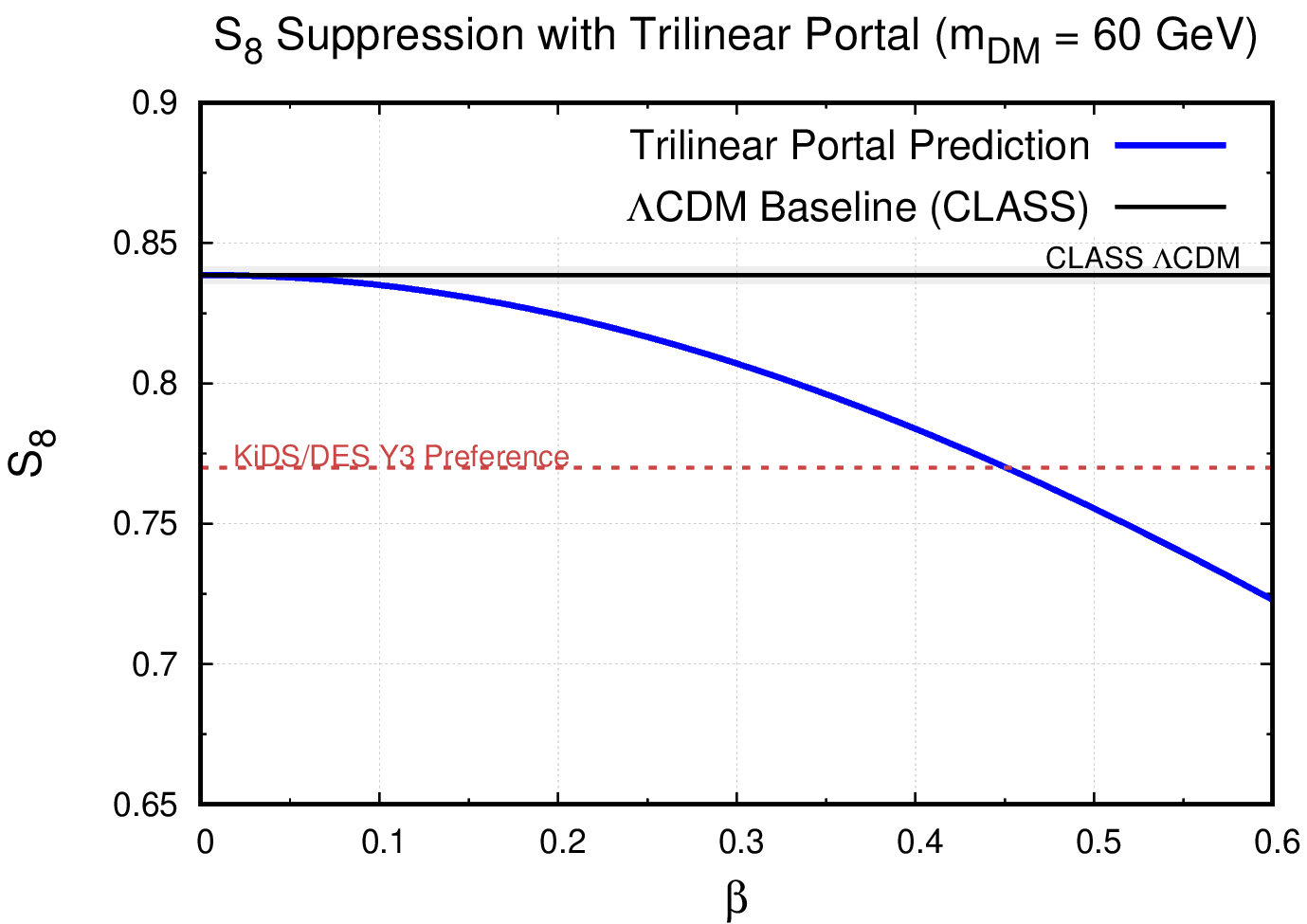}
		\caption{\label{fig:trilinear} $S_8$ suppression as a function of the dimensionless effective coupling $\beta$ for the trilinear portal with $m_\chi=60$~GeV. The $\Lambda$CDM baseline (black) and the KiDS/DES~Y3 preference band (dashed red) are shown for reference.}
	\end{figure}
	
	Figure~\ref{fig:trilinear} shows the late-time structure amplitude $S_8\equiv\sigma_8\sqrt{\Omega_m/0.3}$ as a function of the dimensionless fifth-force parameter $\beta\equiv g_{\rm tri}M_{\rm Pl}/m_\chi^2$ (with $m_\chi=60$~GeV held fixed).  The CLASS
	$\Lambda$CDM baseline sits at $S_8=0.839$; the KiDS-1000 and DES Year~3 preference band lies near $S_8\approx 0.77$ \cite{DESY3}.The trilinear portal suppresses $S_8$ monotonically with increasing$\beta$, crossing the target value at $\beta\approx 0.45$.  Thiscorresponds to a physical Lagrangian coupling $g_{\rm tri}=\beta
	m_\chi^2/M_{\rm Pl}\sim 7\times 10^{-16}$~GeV, confirming the
	phenomenological requirement quoted in Sec.~\ref{sec:trilinear}.
	However, the one-loop Coleman--Weinberg correction to the pNGB mass from the heavy $\chi$ loop imposes the technical-naturalness bound
	$g_{\rm tri}\lesssim 10^{-42}$~GeV (Eq.~\ref{eq:g_natural}). The phenomenological coupling therefore overshoots the naturalness limit by $\sim 26$ orders of magnitude, implying a fine-tuning $\Delta\equiv
	\delta m_\phi^2/m_\phi^2\sim (g_{\rm pheno}/g_{\rm nat})^2\sim
	10^{52}$. The trilinear portal can suppress structure, but only at the cost of destabilizing the ultralight dark-energy mass to the weak scale, thus validating our no-go theorem.
	
	The same CLASS curve (Fig.~\ref{fig:trilinear}) applies identically to the fermionic Yukawa portal (Sec.~\ref{sec:fermion}), because the non-relativistic mapping $g\leftrightarrow y$ and $\chi^2\leftrightarrow\bar\psi\psi$ leaves the dimensionless fifth-force parameter $\beta$ unchanged. For a Majorana WIMP with $m_\psi\sim100$~GeV, the phenomenological coupling required for $S_8\approx0.77$ is $y_{\rm pheno}=\beta\,m_\psi/M_{\rm Pl}\sim2\times10^{-17}$ at $\beta\approx0.45$. The one-loop fermion Coleman--Weinberg correction (Eq.~\ref{eq:delta_mphi_fermion}) imposes the naturalness bound $y\lesssim10^{-43}$ (Eq.~\ref{eq:y_nat}), so the phenomenological coupling overshoots the naturalness limit by $\sim26$ orders of magnitude. The implied tuning is $\Delta\sim(y_{\rm pheno}/y_{\rm nat})^2\sim10^{52}$, identical in order of magnitude to the scalar trilinear case. In a supersymmetric completion (Eq.~\ref{eq:delta_mphi_susy}), the bound relaxes to $y\lesssim10^{-42}$, yielding $\Delta_{\rm SUSY}\sim10^{50}$ (Eq.~\ref{eq:delta_susy}), which is still catastrophically tuned. The fermionic portal therefore reproduces the trilinear no-go precisely.
	
	\begin{figure}
		\includegraphics[width=\columnwidth]{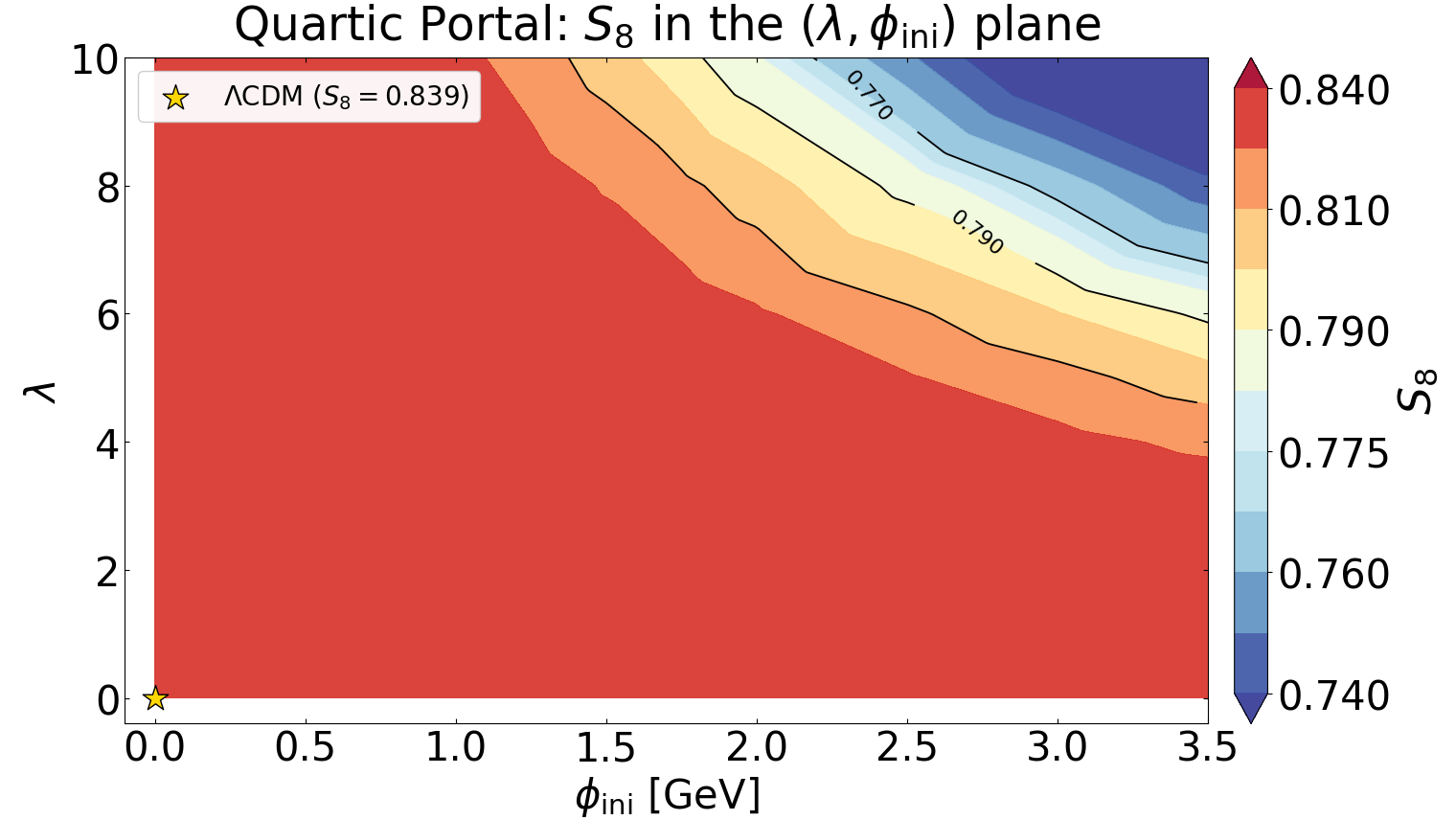}
		\caption{\label{fig:quartic} Contours of constant $S_8$ for the quartic portal with $m_\chi=60$~GeV. Each colored cell corresponds to a direct \texttt{CLASS} computation with linear interpolation between grid points. The $\Lambda$CDM baseline ($S_8=0.839$) is marked by the gold star at the origin. Resolving the $S_8$ tension requires $\lambda\phi_{\rm ini}\sim 20$--$30$ (blue-shaded region), whereas radiative stability demands $\lambda\lesssim 10^{-86}$; even $\phi_{\rm ini}\to M_{\rm Pl}$ would require $\lambda\sim 10^{-18}\gg 10^{-86}$.}
	\end{figure}
	
	Figure~\ref{fig:quartic} displays a two-dimensional scan of $S_8$ in the $(\lambda,\phi_{\rm ini})$ plane, obtained from 208 independent \texttt{CLASS} runs with $m_\chi=60$~GeV and $m_\phi\sim H_0$. The plot reveals an approximate product degeneracy: contours of constant $S_8$ trace hyperbolae $\lambda\,\phi_{\rm ini}={\rm const}$, confirming that the effective fifth-force strength depends only on the combination $\lambda\phi_{\rm ini}/m_\chi^2$. The $\Lambda$CDM point ($\lambda=0$, $\phi_{\rm ini}=0$) anchors the lower-left corner at $S_8=0.839$. Reaching the observed deficit $S_8\approx 0.77$ requires $\lambda\sim 7$--$10$, for $\phi_{\rm ini}\lesssim 4$~GeV. Yet the quartic portal induces a quadratic Coleman--Weinberg divergence $\delta m_\phi^2\sim \lambda m_\chi^2/(16\pi^2)$, yielding the naturalness bound $\lambda\lesssim 10^{-86}$ (Eq.~\ref{eq:lambda_natural}). The phenomenological requirement therefore violates radiative stability by $\sim 87$ orders of magnitude, corresponding to a tuning catastrophe. Even pushing the initial field displacement to the Planck scale, $\phi_{\rm ini}\sim M_{\rm Pl}\sim 10^{19}$~GeV, would still demand $\lambda\sim 10^{-18}\gg 10^{-86}$; the no-go persists because the bound constrains $\lambda$ alone, not the product $\lambda\phi_{\rm ini}$.

	The dimension-6 derivative coupling $(c_6/\Lambda^2)(\partial_\mu\phi)^2\chi^2$ is protected by the shift symmetry $\phi\to\phi+{\rm const.}$ and therefore generates no Coleman--Weinberg correction to $m_\phi$. It is technically natural. However, as demonstrated in the companion work~\cite{Farhan2026}, the momentum-exchange rate $\Gamma$ between dark matter and the ultralight field grows with the DM density and becomes comparable to $H$ at late times ($z\lesssim 2$). Once $\Gamma\gtrsim H$, the two fluids achieve velocity equilibrium and further increases in the coupling do not enhance the suppression of structure growth. This dynamical saturation caps the $S_8$ reduction at $\lesssim 4\%$, insufficient to reconcile the observed $\sim 5$--$10\%$ deficit. No amount of tuning can overcome the saturation limit because it is a purely dynamical, not radiative, obstruction.
	
	The three scalar symmetry-protected portals and the minimal fermionic Yukawa portal exhaust the minimal coupling structures between dark matter and a pNGB dark-energy field. The trilinear and fermionic portals are each radiatively unstable by 26 orders of magnitude, with tuning catastrophes $\Delta\sim10^{52}$ (and $\Delta_{\rm SUSY}\sim10^{50}$ even with supersymmetric cancellation); the quartic portal by 87; and the derivative portal, while radiatively stable, is dynamically saturated at $\lesssim 4\%$ suppression. Consequently, within the class of symmetry-protected scalar and minimal Yukawa portals analyzed here, no single-mediator model simultaneously satisfies technical naturalness ($\Delta<<10^3$) and resolves the $S_8$ tension ($S_8\approx 0.77\pm 0.02$). The impossible triangle of Sec.~\ref{sec:intro} is established within this framework by explicit calculation, and results are summarized in the table below.

	\begin{table}[H]
		\centering
		\renewcommand{\arraystretch}{1.35}
		\setlength{\tabcolsep}{6pt}
		\begin{tabular}{p{3.0cm} p{3.6cm} p{4.4cm} p{2.6cm}}
			\hline\hline
			\textbf{Portal} & \textbf{Naturalness bound} & \textbf{Phenomenological bound} & \textbf{Tuning $\Delta$} \\
			\hline
			Trilinear, $g\phi\chi^2$ & $g\lesssim10^{-42}$~GeV & $g\sim7\times10^{-16}$~GeV  & $\sim10^{52}$ \\[2pt]
			Fermionic, $y\phi\bar{\psi}\psi$ & $y\lesssim10^{-43}$ & $y\sim2\times10^{-17}$  & $\sim10^{52}$ ($\sim10^{50}$ SUSY) \\[2pt]
			Quartic, $\frac{1}{2}\lambda\phi^2\chi^2$ & $\lambda\lesssim10^{-86}$ & $\lambda\sim7$--$10$ ($\phi_{\rm ini}\sim4$~GeV) & $\sim10^{87}$ \\[2pt]
			Derivative, $\frac{c_6}{\Lambda^2}(\partial_\mu\phi)^2\chi^2$ & None & Dynamically saturated: $\Gamma\sim H$; $S_8\lesssim4\%$ & N/A  \\
			\hline\hline
		\end{tabular}
		\caption{Summary of naturalness bounds, phenomenological requirements, and implied fine-tuning for the four DM--DE portals analyzed in this work. The trilinear, fermionic, and quartic portals all violate technical naturalness by many orders of magnitude; the derivative portal is radiatively stable but dynamically saturated, failing to achieve the required $\sim$5--10\% $S_8$ suppression regardless of coupling strength.}
		\label{tab:summary}
	\end{table}

	\section{Multi-field escape routes and their price}
	\label{sec:escape}
	
The no-go theorems of Secs.~\ref{sec:trilinear}--\ref{sec:derivative} assume a single-mediator topology: one pNGB $\phi$ coupled directly to one DM species, whether scalar ($\chi$) or fermionic ($\psi$). One may ask whether multi-field constructions evade the bound by diluting the effective coupling across a chain of $N$ fields. In this section we analyze the minimal such mechanism (the clockwork chain) and prove that the zero-mode tuning cannot be reduced below the single-field value. The clockwork geometrically suppresses the portal, but the radiative correction to the ultralight zero-mode mass is invariant under the gear ratio $q$ and the number of sites $N$.
	
	\subsection*{Standard clockwork with heavy modes at the weak scale}
	
	Consider a clockwork chain of $N$ scalars $\pi_1,\dots,\pi_N$ with nearest-neighbour 
	``gear'' couplings, soft breaking at site~1 to generate the zero-mode mass, and a 
	single site $N$ coupled to dark matter \cite{Giudice2017}. The effective Lagrangian is
	\begin{equation}
		\mathcal{L} \supset \frac{1}{2}\sum_{i=1}^{N}(\partial_\mu\pi_i)^2 
		- \frac{1}{2}m_\pi^2\sum_{i=1}^{N-1}(\pi_i - q\,\pi_{i+1})^2 
		- \frac{1}{2}\mu_{\rm sb}^2\,\pi_1^2
		+ g_{\rm cw}\,\pi_N\,\chi^2 ,
		\label{eq:clockwork_lag}
	\end{equation}
	where $q>1$ is the gear ratio, $m_\pi$ is the heavy-mode scale, and $g_{\rm cw}$ is 
	the bare trilinear coupling at the last site. The mass matrix has $N-1$ heavy modes 
	with $m_{\rm heavy}\sim m_\pi$ and one exponentially light zero mode
	\begin{equation}
		\phi_{\rm eff} = \frac{1}{\mathcal{N}}\sum_{i=1}^{N} q^{-i}\pi_i ,
		\qquad m_{\rm eff}\sim \frac{\mu_{\rm sb}}{q^{N}} \sim H_0 ,
		\label{eq:zero_mode}
	\end{equation}
	where $\mathcal{N}^2 = \sum_i q^{-2i}$. The zero mode couples to DM with an effective 
	coupling suppressed by the clockwork factor
	\begin{equation}
		g_{\rm eff} = c_N\, g_{\rm cw}, \qquad c_N \equiv \frac{q^{-N}}{\mathcal{N}} 
		\sim q^{-N} ,
		\label{eq:geff_cw}
	\end{equation}
	so that $g_{\rm eff} = g_{\rm cw}/q^{N}$ up to $\mathcal{O}(1)$ normalization.
	
	To obtain the phenomenological coupling $g_{\rm eff}\sim 7\times 10^{-16}$~GeV 
	required for $S_8$ resolution (Sec.~\ref{sec:trilinear}), one may choose a much 
	larger bare coupling $g_{\rm cw}\sim\mathcal{O}(1)$~GeV provided 
	$q^{N}\sim 10^{16}$. With a modest gear ratio $q\sim 3$, this requires 
	$N\gtrsim 34$ sites.

	The $\chi$ loop generates a Coleman--Weinberg correction to the mass of site $N$:
	\begin{equation}
		\delta m_N^2 \sim \frac{g_{\rm cw}^2}{8\pi^2}
		\ln\!\left(\frac{\Lambda^2}{m_\chi^2}\right) .
		\label{eq:dm2_clockwork}
	\end{equation}
	Projected onto the $a$-th heavy mode $\psi_a$, this yields a tuning
	\begin{equation}
		\Delta_a \sim \frac{\delta m_a^2}{m_\pi^2}
		\sim \frac{g_{\rm cw}^2}{8\pi^2\,m_\pi^2}
		\ln\!\left(\frac{\Lambda^2}{m_\chi^2}\right).
		\label{eq:heavy_tuning}
	\end{equation}
	For $m_\pi\sim 100$~GeV, $g_{\rm cw}\sim 1$~GeV and $m_\chi\sim 60$~GeV, one finds 
	$\Delta_a \ll 1$; the heavy modes require \emph{no} fine-tuning. 
	Raising $m_\pi$ only improves this. Thus the heavy-mode sector can be perfectly natural.

	Below the scale $m_\pi$, the heavy clockwork modes decouple and the effective theory 
	contains only the zero mode $\phi_{\rm eff}$ with the trilinear portal 
	$g_{\rm eff}\,\phi_{\rm eff}\chi^2$. The Coleman--Weinberg correction to the 
	zero-mode mass is obtained by projecting the site-$N$ correction onto the zero-mode 
	wavefunction:
	\begin{equation}
		\delta m_{\rm eff}^2 = c_N^2\,\delta m_N^2
		\sim q^{-2N}\times\frac{(g_{\rm eff}\,q^{N})^2}{8\pi^2}
		\ln\!\left(\frac{\Lambda^2}{m_\chi^2}\right)
		\label{eq:zero_cw}
	\end{equation}
	
	Thereby, yielding:
	
	$$\delta m_{\rm eff}^2	= \frac{g_{\rm eff}^2}{8\pi^2}
	\ln\!\left(\frac{\Lambda^2}{m_\chi^2}\right).
	$$
	
	The clockwork factors cancel \emph{exactly}: $\delta m_{\rm eff}^2$ depends only on 
	the effective coupling $g_{\rm eff}$, which is fixed by the $S_8$ data. The resulting 
	zero-mode tuning is therefore
	\begin{equation}
		\Delta_0 \equiv \frac{|\delta m_{\rm eff}^2|}{m_{\rm eff}^2}
		\sim \frac{g_{\rm eff}^2}{8\pi^2\,m_{\rm eff}^2}
		\ln\!\left(\frac{\Lambda^2}{m_\chi^2}\right)
		\sim 10^{52},
		\label{eq:delta_zero}
	\end{equation}
	identical to the single-field trilinear portal (Sec.~\ref{sec:trilinear}). The 
	geometric suppression of the portal is undone by the enhanced bare coupling, leaving 
	the radiative correction invariant.
	
	\subsection*{The pNGB clockwork limit}
	
	One may instead insist that \emph{all} clockwork modes are pNGBs with 
	$m_\pi\sim m_{\rm eff}\sim H_0$. In this limit the heavy-mode naturalness bound 
	tightens to $g_{\rm cw}\lesssim 10^{-42}$~GeV (Eq.~\eqref{eq:heavy_tuning} with 
	$m_\pi\sim H_0$). The maximum achievable effective coupling is then
	\begin{equation}
		g_{\rm eff} = \frac{g_{\rm cw}}{q^{N}} \lesssim \frac{10^{-42}\,\mathrm{GeV}}{q^{N}} 
		<< 10^{-42}\,\mathrm{GeV},
	\end{equation}
	which is at least $26$ orders of magnitude below the $g_{\rm eff}\sim 10^{-16}$~GeV 
	required for observable $S_8$ suppression. Consequently, the pNGB clockwork cannot 
	reach the phenomenological target at all, regardless of the number of sites.

	The clockwork construction is the most efficient known multi-field mechanism for 
	suppressing an effective portal coupling. While it renders the heavy modes natural 
	for $m_\pi\gg H_0$, it cannot shield the ultralight zero mode from its own 
	radiative correction. The cancellation in Eq.~\eqref{eq:zero_cw} shows that 
	$\Delta_0\sim 10^{52}$ is a \emph{hard floor}: no choice of $q$, $N$, or $m_\pi$ can 
	reduce the zero-mode tuning below the single-field value. The impossible triangle is therefore a structural boundary within the class of scalar and Yukawa portals and multi-field suppressions analyzed here; it persists even when the single-mediator vertex is replaced by a clockwork chain.
	
	\section{Conclusion}
	\label{sec:conclusion}
	
	We have presented a systematic no-go analysis of interacting dark energy (IDE) in the context of the $S_8$ tension. Anchored in the $Z_2$-symmetric Inert Doublet + Complex Singlet Model (Sec.~\ref{sec:framework}), a renormalizable UV completion with a pNGB dark-energy field and a WIMP dark-matter candidate, we examined three minimal symmetry-protected scalar portals between the dark sectors: trilinear ($g\phi\chi^2$), quartic ($\lambda\phi^2\chi^2$), and derivative ($(c_6/\Lambda^2)(\partial\phi)^2\chi^2$). Each portal was studied in isolation to isolate its individual no-go constraint, and the trilinear and quartic cases were validated numerically with a modified version of the \texttt{CLASS} Boltzmann code.
	
	For the trilinear portal, numerical integration shows that resolving the $S_8$ tension ($S_8\approx 0.77$) requires a dimensionless effective coupling $\beta \sim 0.45$, i.e.~$g\sim 10^{-16}$~GeV for $m_\chi=60$~GeV (Fig.~\ref{fig:trilinear}). This exceeds the one-loop Coleman--Weinberg naturalness bound $g\lesssim 10^{-42}$~GeV by $\sim 26$ orders of magnitude, implying a fine-tuning $\Delta\sim 10^{52}$. The quartic portal exhibits an approximate product degeneracy $\lambda\phi_{\rm ini}={\rm const.}$, reaching $S_8\approx 0.77$ requires $\lambda\phi_{\rm ini}\sim 20$--$30$ (Fig.~\ref{fig:quartic}), which for any initial displacement demands $\lambda\gg 10^{-86}$. The resulting tuning is even more catastrophic. The derivative portal, protected by the pNGB shift symmetry, is technically natural but dynamically saturates: momentum-exchange drives the dark sectors to velocity equilibrium at late times, capping structure suppression at $\lesssim 4\%$, which is insufficient to reconcile the $\sim$5--10\% observed deficit.
	
    We extended the analysis to the minimal fermionic Yukawa portal $y\phi\bar{\psi}\psi$ (Sec.~\ref{sec:fermion}). In the non-relativistic limit the cosmological phenomenology maps identically onto the scalar trilinear portal, so the same $\beta\approx0.45$ is required for $S_8$ suppression. The one-loop fermion Coleman--Weinberg correction enforces $y\lesssim10^{-43}$, while the phenomenological target is $y\sim10^{-17}$, yielding $\Delta\sim10^{52}$. In a supersymmetric completion the quadratic divergences cancel between the fermion and its scalar partner, but the soft-breaking scale $M_{\rm soft}\sim m_{\psi}$ for viable thermal relics means the tuning floor remains $\Delta_{\rm SUSY}\sim10^{50}$. Consequently, the impossible triangle applies to both scalar and fermionic dark matter, with and without supersymmetric cancellation.
	
	We also examined the minimal multi-field escape, a clockwork chain of $N$ scalars (Sec.~\ref{sec:escape}). In the standard implementation with heavy modes at the weak scale ($m_\pi\sim 100$~GeV), the bare coupling can be natural ($g_{\rm cw}\sim\mathcal{O}(1)$~GeV) and the heavy modes require no tuning; however, the radiative correction to the ultralight zero-mode mass depends only on the effective coupling, $\delta m_{\rm eff}^2\propto (c_N g_{\rm cw})^2 = g_{\rm eff}^2$, so the geometric clockwork suppression cancels out exactly. Consequently the zero-mode tuning remains $\Delta_0\sim 10^{52}$, identical to the single-field case, a \emph{hard floor}, that no choice of $q$ or $N$ can reduce. In the conservative limit where all modes are pNGBs with $m_\pi\sim H_0$, the heavy-mode bound tightens to $g_{\rm cw}\lesssim 10^{-42}$~GeV, making the phenomenological target $g_{\rm eff}\sim 10^{-16}$~GeV unreachable regardless of the number of sites. In either regime, the impossible triangle persists.
	
	The impossible triangle of (i)~radiative stability, (ii)~successful $S_8$ resolution, and (iii)~single-mediator topology is a structural boundary for the scalar-mediated and Yukawa-mediated dark-sector portals analyzed here. Within this framework, it persists across the three minimal symmetry-protected scalar couplings, the fermionic Yukawa coupling, and survives even the most efficient known multi-field suppression mechanism. Resolving the $S_8$ tension through IDE requires either abandoning technical naturalness with explicit, quantified fine-tuning, or invoking a yet-unknown symmetry or dynamical mechanism that operates beyond the portal framework analyzed here. We emphasize that our no-go theorem applies specifically to the class of perturbative, single-mediator scalar and Yukawa portals analyzed here. It does not exclude nonperturbative screening mechanisms~\cite{Mota2006}, vector dark matter, modified gravity embeddings, or multi-field constructions beyond the clockwork chain. Identifying whether any of these alternatives can simultaneously achieve naturalness and $S_8$ suppression remains an open direction. Our results map the precise price of each alternative, providing a reference point for future dark-sector constructions.

\end{document}